\documentclass[a4paper,nofootinbib,preprintnumbers,twocolumn,preprintnumbers,floatfix,superscriptaddress,prl,showpacs]{revtex4-1}

\usepackage{graphicx}
\usepackage{amssymb}
\usepackage{amsmath}
\usepackage{epstopdf}
\usepackage{graphics}
\usepackage[caption=false]{subfig}
\usepackage{float}
\usepackage{color}
\usepackage{hyperref}

\allowdisplaybreaks[1]

\newcommand{\avg}[1]{\langle #1 \rangle} 

\newcommand{\figref}[1]{Fig.~\ref{#1}}

\begin{document}

\title{Efficient atomic clocks operated with several atomic ensembles}

\author{J. Borregaard }
\author{A. S. S\o rensen}

\affiliation{QUANTOP, The Niels Bohr Institute, University of Copenhagen, Blegdamsvej 17, DK-2100 Copenhagen \O, Denmark}

\date{\today}

\begin{abstract}
Atomic clocks are typically operated by locking a local oscillator (LO) to a single atomic ensemble. In this article we propose a scheme where the LO is locked to several atomic ensembles instead of one. This results in an exponential improvement compared to the conventional method and provides a stability of the clock scaling as $(\alpha N)^{-m/2}$ with $N$ being the number of atoms in each of the $m$ ensembles and $\alpha$ is a constant depending on the protocol being used to lock the LO.   
\end{abstract}


\maketitle

Atomic clocks provides very precise time measurements useful for a broad range of areas in physics. The quantum noise of the atoms limits the stability of atomic clocks, resulting in the standard quantum limit where the stability scales as  $1/\sqrt{N}$ with $N$ being the number of atoms \cite{Santarelli1999,itano1993pra}. Various ways of improving the resolution have been suggested such as using entangled states with reduced atomic noise \cite{wineland1994pra,bollinger1996pra,andre2004prl,lloydnature2011} to push the resolution to the Heisenberg limit where it scales as $1/N$ \cite{leibfried2004science,Leroux,Appel,Anne,riedel,gross}. Another approach to increasing the stability is to use optical atomic clocks where the higher operating frequency leads to an improved stability \cite{winelandscience2001,udemnature2002,margolisscience2004,takamotonature2005,rosenbandscience2008}. Since an atomic clock is typically operated through Ramsey spectroscopy~\cite{ramsey1956} the resolution can also be enhanced by increasing the Ramsey time $T$ resulting in an improvement scaling as $1/\sqrt{T}$ \cite{wineland1998prl,wineland1998jrnist,laurantepjd1998}. For clocks with trapped atoms, where there are no other limitations, $T$ becomes limited only by the decoherence in the system. In practice this decoherence often originates from the frequency fluctuations of the local oscillator (LO) used to drive the atomic clock transition~\cite{wineland1998jrnist}. Hence, the stability can also be increased by simply devising methods to increase the Ramsey period by stabilizing the LO \cite{shiganjp2012}. 

\begin{figure} 
\centering
\subfloat {\label{fig:schsd}\includegraphics[width=0.22\textwidth]{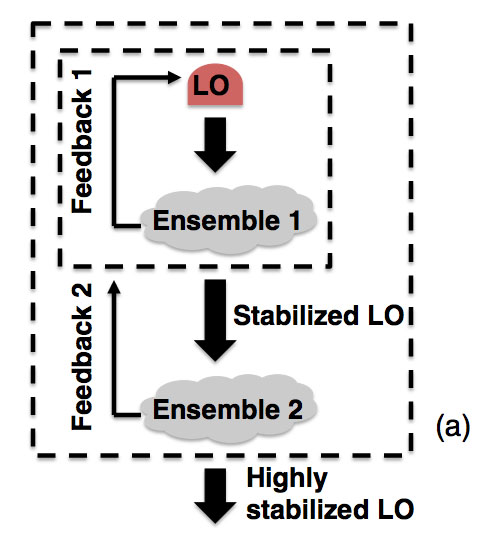}} 
\subfloat{\label{fig:w_feedback}\includegraphics[width=0.28\textwidth]{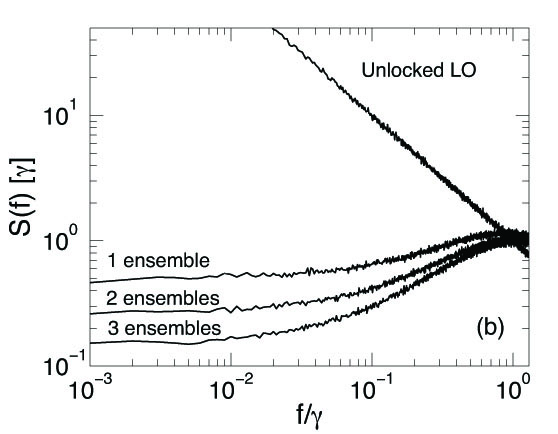}}
\caption{(Colour online) (a) Illustration of locking the LO using several ensembles. The feedback of the first ensemble stabilizes the LO such that the second ensemble can be operated with a longer Ramsey time. The feedback from the second ensemble then further stabilizes the LO. (b) Numerical simulation of the frequency noise spectrum $S(f)$ of the LO when locked to between 1 and 3 ensembles. The data was simulated as described in the text for $N=20$ and $T_{1}=0.1/\gamma$ for the conventional Ramsey scheme. The first feedback lowers the noise of the LO and whitens the spectrum even though the unlocked LO was assumed to be subject to $1/f$ noise. The second and third feedbacks further lowers the noise of LO by a constant factor.}
\label{fig:scheme}
\end{figure}

In this letter we suggest a scheme where the frequency of the LO is locked to the atomic transition using several ensembles of atoms. This procedure allows increasing the Ramsey period each time another ensemble is used. As a result we find that the stability of the clock can increase exponentially with the number of ensembles. \figref{fig:scheme}(a) illustrates the idea behind the scheme. 
The feedback of the first ensemble locks the frequency of the LO thus reducing the noise to the atomic noise. Having reduced the noise in the LO, the second ensemble can be operated with a longer Ramsey time. Through a second feedback the noise of the LO can be further reduced as shown in the simulation in \figref{fig:scheme}(b) (details are given later). The procedure can be extended to any number of ensembles and for uncorrelated atoms the stability of the LO will scale as $\sqrt{\gamma}(\gamma T_{1}N)^{-m/2}$ where $m$ is the number of ensembles (each containing $N$ atoms), $\gamma$ is a parameter characterizing the frequency fluctuations of the unlocked LO, and $T_{1}$ is the Ramsey time of the first ensemble. Hence the scheme can provide an exponential improvement in the stability with the total number of atoms. In order for the clock to be stable we need $\gamma T_{1} \ll 1$ and hence the protocol requires a minimum number of atoms to improve the performance. With the conventional Ramsey protocol we find that the scheme works for a minimum ensemble size of $20$ atoms. To further optimize the performance of the scheme we study an adaptive measurement protocol for estimating the LO frequency offset, which extends the applicability of the scheme down to ensembles with only 4 (7) atoms for white ($1/f$) noise in the LO. This makes the scheme relevant for atomic clocks based on trapped ions, which are typically constructed with only a few ions \cite{wineland1998prl}. A related procedure involving multiple measurements on a single ensemble was proposed in Ref. \cite{shiganjp2012}. By using multiple ensembles our procedure avoids disturbances from the measurements affecting later measurements. Recently and independently from this work a manuscript appeared, which treats essentially same locking scheme that we suggest \cite{rosenbandarxiv2013}. Taking the different figures of merit into account that work arrives at results consistent with ours.    

We will now describe the locking of the LO to the atomic transition using Ramsey spectroscopy. We model an ensemble of $N$ atoms as a collection of spin-1/2 particles with total angular momentum $\vec{J}$. We define the angular momentum operators $\hat{J}_{x},\hat{J}_{y}$ and $\hat{J}_{z}$ in the usual way and initially the atoms are pumped to have $\avg{\vec{J}}$ along the $z$-direction, $\avg{\hat{J}_{x}}=\avg{\hat{J}_{y}}=0$. In Ramsey spectroscopy the atoms are illuminated by a near-resonant $\pi/2$-pulse from the LO, followed by the Ramsey time $T$ of free evolution, and finally another near-resonant $\pi/2$-pulse is applied. The Heisenberg evolution of $\hat{J}_{z}$ is $\hat{J}_{3}=\cos(\delta\phi)\hat{J}_{y}+\sin(\delta\phi)\hat{J}_{z}$ where $\delta\phi=\delta\omega� T$ is the acquired phase of the LO relative to the atoms. At the end of the Ramsey sequence $\hat{J}_{3}$ is measured  and used to make an estimate $\delta\phi^{e}=\!-\!\arcsin(2\hat{J}_{3}/N)$ of $\delta\phi$.  The feedback loop then steers the frequency of the LO towards the atomic transition by applying a  frequency correction of $\Delta\omega=-\alpha \delta\phi^{e}/T$ to the LO where $\alpha$ sets the strength of the feedback loop. The operation of an atomic clock thus consists of repeating a cycle of initializing - Ramsey sequence - measurement - feedback. The total time of this clock cycle is denoted $T_{c}$ and we assume that  $T_{c}\sim T$, i.e. we assume a negligible Dick noise~\cite{dick}.

We now consider an atomic clock with two atomic ensembles operated with different Ramsey times and show how this can improve the stability of the clock. These considerations can then easily be extended to several ensembles. Note that we assume the intrinsic linewidth of the atoms to be negligible such that the atomic linewidth is only limited by the Ramsey time. The first ensemble is operated with Ramsey time $T_{1}$ and we assume that the second ensemble is operated with Ramsey time $T_{2}=nT_{1}$ where $n$ is an integer. We can make two discrete time scales describing ensemble one and two respectively. Ensemble one is measured at $t_{k}=kT_{1}$ and ensemble two is measured at $t_{s}=sT_{2}=s\cdot nT_{1}$. The frequency offset of the LO between time $t_{k\!-\!1}$ and $t_{k}$ is then 
\begin{equation} \label{eq:freqoff1}
\delta\omega(t)=\delta\omega_{0}(t)+\Delta\omega_{1}(t_{k-1})+\Delta\omega_{2}(t_{s-1}),
\end{equation}
where $\delta\omega_{0}(t)$ is the frequency fluctuation of the unlocked LO, $\Delta\omega_{1}(t_{k-1})$ is the sum of the frequency corrections applied up to time $t_{k-1}$ from the first ensemble and $\Delta\omega_{2}(t_{s-1})$ is the sum of the frequency corrections applied up to time $t_{s\!-\!1}$ from the second ensemble ($t_{s-1}\leq t_{k-1}$). The feedback loops are described by the equations  
\begin{eqnarray}
\Delta\omega_{1}(t_{k-1})&=&\Delta\omega_{1}(t_{k-2})-\alpha\delta\phi^{e_{1}}(t_{k-1})/T_{1} \label{eq:correct1} \\
\Delta\omega_{2}(t_{s-1})&=&\Delta\omega_{2}(t_{s-2})-\alpha\delta\phi^{e_{2}}(t_{s-1})/T_{2}, \label{eq:correct2}
\end{eqnarray}
where $\delta\phi^{e_{1}}(t_{k-1})$ and $\delta\phi^{e_{2}}(t_{s-1})$ are the estimated phases from the first and second ensemble at times $t_{k-1}$ and $t_{s-1}$ respectively. Using Eq. \eqref{eq:freqoff1} we can write the phase of the LO relative to the atoms of the second ensemble at time $t_{s}$ as
\begin{eqnarray} \label{eq:phase2}
 \delta\phi_{2}(t_{s})=\int_{0}^{T_{2}}\!\!\!\!\!\!\! \text{d}t'\delta\omega(t_{s}\!-\!t')=\Delta\phi_{s-1} + \delta\tilde{\phi}(t_{s}),
 \end{eqnarray}
 where $\Delta\phi_{s-1}\!=\!\!\int_{0}^{T_{2}}\Delta\omega_{2}(t_{s-1})\text{d}t'$ is the accumulated phase due to the feedback of the second ensemble and 
 \begin{eqnarray} \label{eq:stabphase1}
 \delta\tilde{\phi}(t_{s})\!=\!\!\int_{0}^{T_{2}}\!\!\!\!\!\!\! \text{d}t' \delta\tilde{\omega}(t_{s}\!-\!t')\!=\!\!\!\int_{0}^{T_{2}}\!\!\!\!\!\!\! \text{d}t'\delta\omega_{0}(t_{s}\!-\!t')\!+\!\Delta\omega_{1}(t_{s}\!-\!t')\quad
\end{eqnarray}
is the accumulated phase due to the frequency oscillations of the LO when locked by the feedback of the first ensemble. For now we assume that $T_{2}\gg T_{1}$ such that the feedback of the first ensemble has stabilized the LO but later we will relax this assumption. From Eqs. \eqref{eq:correct2}-\eqref{eq:phase2} we then derive the difference equation
 \begin{eqnarray} \label{eq:diffeq}
 \delta\phi_{2}(t_{s})\!-\!\delta\phi_{2}(t_{s-1})=\delta\tilde{\phi}(t_{s})\!-\!\delta\tilde{\phi}(t_{s-1})\!-\!\alpha\delta\phi^{e_{2}}(t_{s-1}).\quad
 \end{eqnarray}
 From this expression we see that the evolution of the second phase $\delta\phi_{2}$ is essentially driven by the noise of the stabilized LO from the first step $\delta\tilde{\phi}$ but is stabilized by the second feedback loop described by $\alpha\delta\phi^{e_{2}}$.
 
To solve Eq. \eqref{eq:diffeq} we need to characterize the width of the noise of the stabilized LO from the first stage, $\avg{\delta\tilde{\phi}^{2}}=\int_{0}^{T_{2}}\!\!\text{d}t\int_{0}^{T_{2}}\!\!\text{d}t'\avg{\delta\tilde{\omega}(t)\delta\tilde{\omega}(t')}$. From Eq. \eqref{eq:correct1} and \eqref{eq:stabphase1} we can derive a difference equation for $\delta\tilde{\phi}(t_{k})=\!\int_{0}^{T_{1}}\delta\tilde{\omega}(t_{k}\!-\!t')\text{d}t'$, which is the acquired phase of the LO relative to the first ensemble between time $t_{k-1}$ and $t_{k}$ (we can neglect the feedback from the second ensemble since $T_{2} \gg T_{1}$):
\begin{eqnarray} \label{eq:diffeq2}
\delta\tilde{\phi}(t_{k})\!-\!\delta\tilde{\phi}(t_{k-1})=\delta\phi_{0}(t_{k})\!-\!\delta\phi_{0}(t_{k-1})\!-\!\alpha\delta\phi^{e_{1}}(t_{k-1}).\quad
\end{eqnarray}
Here $\delta\phi_{0}(t_{k})=\!\int_{0}^{T_{1}}\delta\omega_{0}(t_{k}-t')\text{d}t'$ is the phase of the unlocked LO. In comparison to Eq.\eqref{eq:diffeq} we see that the evolution of the phase $\delta\tilde{\phi}$ is driven by the noise of the unlocked LO but is stabilized by the first feedback loop described by $\alpha\delta\phi^{e_{1}}$. To solve this equation we follow Ref.~\cite{andrephd} where the locking of the LO to a single ensemble is described. First we derive a differential equation from Eq. \eqref{eq:diffeq2} in the limit $N\gg1$, treating $\hat{J}_{x},\hat{J}_{y}$, and $\hat{J}_{z}$ as Gaussian variables and considering for now a LO subject to white noise. Assuming that the atoms start out in a coherent spin state we can solve this equation to obtain   
\begin{equation} \label{eq:noisestab}
\avg{\delta\tilde{\phi}^{2}}=T_{2}/NT_{1}=\tilde{\gamma}T_{2},
\end{equation}
where we have defined the parameter $\tilde{\gamma}=1/NT_{1}$, which characterizes the noise of the stabilized LO. This noise is effectively white for both white and $1/f$ noise in the unlocked LO (\figref{fig:scheme}(b) and Ref.~\cite{andrephd}). The second ensemble thus sees an effective white noise in the LO with $\tilde{\gamma}=1/(T_{1}N)$. 

We now return to Eq. \eqref{eq:diffeq}. Writing $\delta\phi_{2}(t)\sim\delta\omega(t)T_{2}$ the stability of the clock after running for a time $\tau \gg T_{2}$ is 
 \begin{eqnarray}
  \sigma_{\gamma}(\tau)\!\!=\!\!\frac{1}{\omega\tau T_{2}}\!\!\left(\int_{0}^{\tau}\!\!\!\!\!\text{d}t\!\!\!\int_{0}^{\tau}\!\!\!\!\!\text{d}t'\avg{\delta\phi_{2}(t)\delta\phi_{2}(t')}\right)^\frac{1}{2}\!\!\!\!\!, \quad  \label{eq:stability1}
 \end{eqnarray}
where $\omega$ is the frequency of the atomic transition. Following similar arguments as before we can derive and solve a differential equation from Eq. \eqref{eq:diffeq} to obtain an expresion for $\avg{\delta\phi_{2}(t)\delta\phi_{2}(t')}$. Inserting this into Eq. \eqref{eq:stability1} and taking the limit of $\tau\gg T_{2}$ results in 
\begin{equation} \label{eq:stability2}
\sigma_{\gamma}(\tau)=\frac{1}{\omega}\sqrt{\frac{1}{\tau NT_{2}}}.
\end{equation}
Eq. \eqref{eq:stability2} describes how the stability improves with $T_{2}$ and $N$. The longest $T_{2}$ we can allow is determined by how well the LO is stabilized by the first ensemble as contained in $\tilde{\gamma}$ and we parameterize it by $T_{2,max}=\beta_{2}/\tilde{\gamma}$. In a similar fashion we assume that $T_{1,max}=\beta_{1}/\gamma$ for the first ensemble. With these parameterizations we can express the stability as 
\begin{equation} \label{eq:stability3}
\sigma_{\gamma}(\tau)=\frac{1}{\omega}\sqrt{\frac{\gamma}{\tau N^{2}\beta_{1}\beta_{2}}}=\frac{1}{\omega}\sqrt{\frac{\gamma\beta_{1}/\beta_{2}}{\tau (N\gamma T_{1,max})^{2}}}.
\end{equation} 
With white noise in the unlocked LO we can pick $\beta_{1}=\beta_{2}$. As previously noted the noise of the LO will also be approximately white with $\tilde{\gamma}\sim1/NT_{1}$ after locking it to the first ensemble also for other types of noise e.g. $1/f$ noise. In that case it is desirable to have $\beta_{2}\neq \beta_{1}$ but we still expect $\beta_{1}/\beta_{2}$ to be of order unity.  Eq. \eqref{eq:stability3} shows that by locking the LO to two ensembles of uncorrelated atoms the stability can be significantly improved. If $N\gamma T_{1} \gg1$ the stability obtained from Eq. \eqref{eq:stability3} is much better than the single ensemble result in Eq. \eqref{eq:stability2} (with $T_{2}\!\to\! T_{1}$). The arguments leading to Eq. \eqref{eq:stability3} can be generalized in a straight forward way to show that if the LO is locked to $m$ ensembles each containing $N$ atoms, the stability of the clock is $\sigma_{\gamma}(\tau)=\sqrt{(\beta_{1}/\beta)^{(m\!-\!1)}\gamma/(\omega^{2}\tau)}(N\gamma T_{1,max})^{-m/2}$ (since the noise of the LO is white after locking it to the first ensemble we use $\beta=\beta_{2}\!=\!\ldots\!=\!\beta_{m}$). By continuing the procedure we thus improve the stability exponentially! 
 
 In our analytical calculations above we have assumed $N\gg1$. To investigate the performance for smaller $N$ we simulate an atomic clock locked to between 1 and 4 atomic ensembles each with atom numbers from $N=20$ to $N=100$. From the simulations we can generalize to the case where the LO is locked to $m$ ensembles. We simulate the full quantum evolution of the atomic state through the Ramsey sequences and subsequent measurements and implement the feedback on the LO similar to the description in Eq. \eqref{eq:freqoff1} and above. The assumption of $T_{2}\gg T_{1}$ can be relaxed by applying a phase correction in the measurement \cite{SM}. The number of atoms required in each ensemble to increase the Ramsey time by a factor $a$ at each level is set by the white noise level of the stabilized LO. Using Eq. \eqref{eq:noisestab} and remembering that $\beta$ parameterize the maximal Ramsey time for white noise we have that $T_{2}/NT_{1}=\tilde{\gamma}T_{2}=\beta$.  Assuming $T_{2}=aT_{1}$ we find that $N\sim a/\beta$ atoms are required in each ensemble to increase the Ramsey time by a factor of $a$ at each level. The minimum number of atoms required for our protocol to work is thus obtained by setting $a=2$. 

To determine $\beta$ we investigate the errors that limits the Ramsey time $T$ for a LO subject to white noise characterized by $\gamma$. For experiments or simulations running with a fixed Ramsey time there is always a finite probability that phase jumps large enough to spoil the measurement strategy occurs since Ramsey spectroscopy with projective measurements is only effective for phases $\lesssim\pi/2$. In our simulations we see these phase jumps as an abrupt break down as we increase $T$. Simulating a clock running for a time $\tau=10^{6}T$ with a single ensemble of $N=10^{5}$ atoms, we see the stability increase with $T$ until a maximum of $T_{max}\sim0.1/\gamma$ is reached. Increasing $T$ beyond this point results in a rapid decrease in the stability. From this we conclude that Ramsey spectroscopy with projective measurements only allows for $\beta\sim0.1$ and thus $N_{min}=20$.  To determine $\beta_{1}$ for an LO subject to $1/f$ noise we do a similar simulation where the noise spectrum of the LO is $S(f)=\gamma^{2}/f$ ($f$ is frequency). From this simulation we find that $\beta_{1}\sim0.1$ as for white noise. Note that this construction introduces a weak (logarithmic) dependence on the number of steps that we simulate \cite{SM}.  

\begin{figure} 
\centering
\subfloat {\label{fig:white}\includegraphics[width=0.25\textwidth]{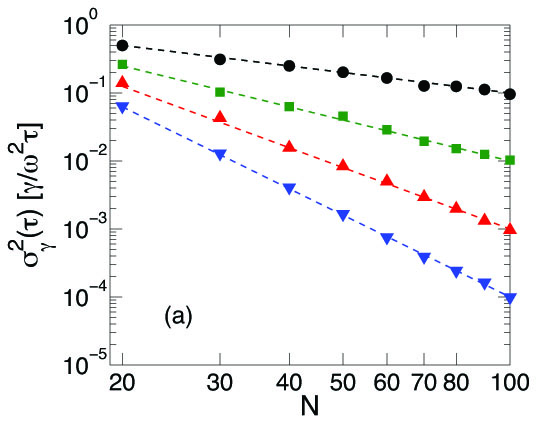}} 
\subfloat{\label{fig:pink}\includegraphics[width=0.25\textwidth]{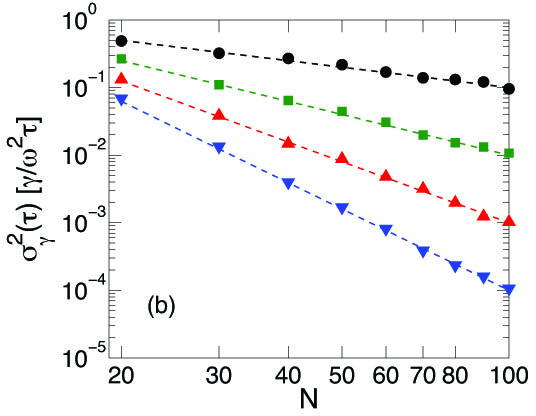}}
\caption{(Colour online) The stability of atomic clocks for a LO subject to (a) white noise and (b) $1/f$-noise. $\bullet,\blacksquare,\blacktriangle$, and $\blacktriangledown$ is the stability of a clock with the LO locked to 1,2,3, and 4 ensembles containing $N$ atoms each. The clocks were simulated with $\beta_{1}=0.1$ and $T_{j}=nT_{j-1}$. Counting from the left (low $N$) the points are for integers $n$ from $2$ to $10$. The dashed lines are the analytical calculations.}
\label{fig:wp}
\end{figure}

We have simulated clocks with an unlocked LO subject to both white and $1/f$ noise with the constraint $\beta=0.1$. In \figref{fig:wp} the stability of the clocks are plotted against the ensemble size $N$. 
\figref{fig:wp} confirms that the scheme works down to atom numbers of $N=20$ where we gain a factor of $\sim 2^{m-1}$ in $\sigma_{\gamma}^{2}(\tau)$ by locking the LO to $m$ ensembles for both white and $1/f$ noise. Furthermore the numerical results are seen to agree nicely with the analytical calculations. We obtain practically the same long term stability for $1/f$ noise as for white noise since the first feedback whitens the noise for small frequencies (cf. \figref{fig:w_feedback}). 

\begin{figure} 
\centering
\subfloat {\label{fig:white}\includegraphics[width=0.25\textwidth]{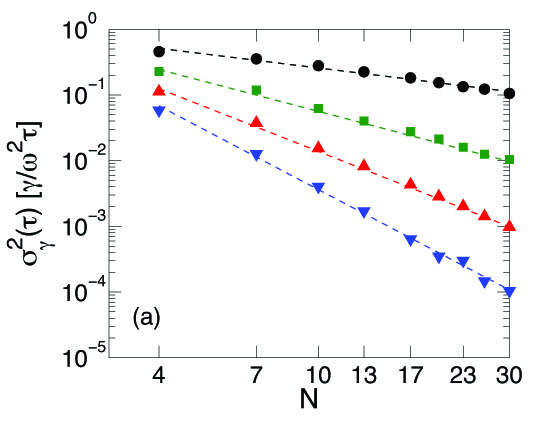}} 
\subfloat{\label{fig:pink}\includegraphics[width=0.25\textwidth]{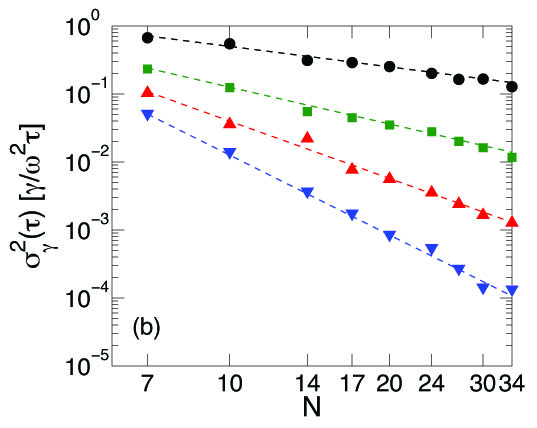}}
\caption{(Colour online) The stability of atomic clocks with adaptive measurements for a LO subject to (a) white noise and (b) $1/f$-noise. $\bullet,\blacksquare,\blacktriangle$, and $\blacktriangledown$ is the stability of a clock with the LO locked to 1,2,3, and 4 ensembles of $N$ atoms each. The adaptive protocol allows for $\beta_{1}=0.3$ (a) and $\beta_{1}=0.2$ (b). The clocks were simulated with $T_{j}=nT_{j-1}$ and counting from the left (low $N$) the points are for integers $n$ from $2$ to $10$. The dashed lines are fits of the simulated data.}
\label{fig:wa}
\end{figure}

The conventional Ramsey protocol considered so far has a lower limit of $N_{min}=20$ in order for our protocol to work. This limit is due to the inability of the conventional protocol to effectively resolve phases larger than $\pi/2$. In Ref.~\cite{johannesarxiv} we presented an adaptive protocol for estimating the phase, which effectively resolves phases $\lesssim\pi$. Again simulating a clock  running for a time $\tau=10^{6}T_{1}$ with a single ensemble of $N=10^{5}$ atoms we find that this protocol enables us to extend the Ramsey time to $\beta\sim0.3$ for white noise and to $\beta_{1}\sim 0.2$ for a LO subject to $1/f$ noise \cite{SM}. However the type of weak measurements described in Ref.~\cite{johannesarxiv} is hard to implement for ensembles of few atoms. We have therefore modified the protocol such that individual atoms are read out one at a time and a Bayesian procedure similar to that of Ref.~\cite{wiseman2000prl,wisemannature2007} is used for the phase estimation and atomic feedback.  We perform intermediate feedbacks during the measurements to rotate the atomic state to be almost in phase with the LO. Due to the rotations the protocol can resolve phases $\lesssim\pi$ as the protocol in Ref.~\cite{johannesarxiv}. This protocol is described in detail in the supplemental material \cite{SM}. With this adaptive measurement strategy we simulate clocks locked to between 1 and 4 ensembles for atom numbers from $N=4$ to 34 with an unlocked LO subject to both white and $1/f$ noise with the constraint $\beta=0.3$. The stability of the clocks is plotted against the ensemble size $N$ in \figref{fig:wa}. 
For the adaptive protocol we can apply the scheme of locking to several ensembles down to ensemble sizes of $N=4$ $(7)$ for white ($1/f$) noise where we gain a factor of $\sim 2^{m-1}$ in $\sigma_{\gamma}^{2}(\tau)$ by locking the LO to $m$ ensembles. The minimal number of atoms is higher for $1/f$ noise since the adaptive protocol is not as effective as for white noise where we have a better understanding of the a priori distribution in the Bayesian procedure~\cite{SM}. It should be noted, however, that in principle it is only in the first ensemble that we need more atoms than for white noise since the feedback of the first ensemble whitens the noise. The adaptive protocol is thus more effective for the subsequent ensembles.   

In conclusion we have demonstrated a scheme for locking the LO in an atomic clock to $m$ ensembles of $N$ atoms each.  For this scheme the stability of the clock scales as $\sqrt{\gamma}(\gamma T_{1}N)^{-m/2}$ where $T_{1}$ is the Ramsey time of the first ensemble. Our scheme thus provide an exponential improvement in the stability with the number of atoms. For the conventional Ramsey protocol our scheme is applicable down to ensemble sizes of $N=20$ atoms while it is applicable down to ensemble sizes of $N=4$ $(7)$ using an adaptive protocol. This make the scheme relevant for atomic clocks with trapped ions. The performance of the protocol can be improved further by considering squeezed states but this is beyond the scope of this article.   

We gratefully acknowledge the support of the Lundbeck Foundation and the Danish National Research Foundation through QUANTOP. The research leading to these results has received funding from the European Research Council under the European Union's Seventh Framework Programme (FP/2007-2013) / ERC Grant Agreement n. 306576. We also thank  D. Wineland, A. Andr\'e and M. Lukin for helpful discussions.

\newpage
\renewcommand{\theequation}{S\arabic{equation}}
\renewcommand{\thefigure}{S\arabic{figure}}
\onecolumngrid

\section{Supplemental material: Efficient atomic clocks operated with several atomic ensembles}

This supplemental material to our article "Efficient atomic clocks operated with several atomic ensembles" describes the details of our numerical simulations of atomic clocks locked to several atomic ensembles and the details of the modified adaptive protocol where the atoms are read out one at a time. Furthermore we show how the assumption of $T_{2} \gg T_{1}$ made in the article  can be relaxed by applying a phase correction in the measurement of the second ensemble and how we find the limit of the free evolution time.

\section{Phase corrections}
 
The Ramsey sequence and the subsequent estimate of the drifted phase of the LO relative to an ensemble of atoms is described in the article. Eq. (1) - (3) in the article describes the frequency offset of the LO ($\delta\omega(t)$) between time $t_{k\!-\!1}=(k\!-\!1)T$ and $t_{k}=kT$ when the LO is locked to two ensembles. We will now generalize this formalism to the case where the LO is locked to $m$ ensembles.  Assuming that the $j$'th ensemble is operated with Ramsey time $T_{j}=n^{j-1}T_{1}$ ($n$ is an integer describing how many times the Ramsey time can be increased for each added ensemble) the frequency offset of the LO between time $t_{k\!-\!1}=(k\!-\!1)T_{1}$ and $t_{k}=kT_{1}$ is
\begin{equation} \label{eq:freqoff}
 \delta\omega(t)=\delta\omega_{0}(t)+\Delta\omega_{1}(t_{s_{1}})+\Delta\omega_{2}(t_{s_{2}n})+\ldots+\Delta\omega_{m}(t_{s_{m}n^{m-1}}),
\end{equation} 
where $\delta\omega_{0}(t_{k})$ is the frequency fluctuations of the unlocked LO and $\Delta\omega_{m}(t_{s_{j}n^{j-1}})$ is the sum of the frequency corrections applied up to time $t_{s_{j}n^{j-1}}$ from the $j$'th ensemble ($s_{j}$ is found by rounding $(k\!-\!1)/n^{j-1}$ down to the nearest integer). Note that the index $s_{j}n^{j-1}$ should be read as $s_{j}$ times $n^{j-1}$ describing the exponential increase in the Ramsey time each time another ensemble is used. The iterative equation for $\Delta\omega_{j}(t_{s_{j}n^{j-1}})$ is
\begin{equation} \label{eq:feedback}
\Delta\omega_{j}(t_{s_{j}n^{j-1}})=\Delta\omega_{j}(t_{(s_{j}-1)n^{j-1}})-\alpha\delta\phi^{e_{j}}(t_{s_{j}n^{j-1}})/T_{j},
\end{equation}
where $\delta\phi^{e_{j}}(t_{s_{j}n^{j-1}})$ is the estimated phase from the $j$'th ensemble at time $t_{s_{j}n^{j-1}}$ and $\alpha$ sets the strength of the feedback loop (for now we assume equal strengths for all feedback loops). $\alpha$ determines how long time the clocks needs to run before the LO is effectively locked by the feedbacks (The LO is locked after a time $\sim T_{j}/\alpha$). In the article we assumed that $T_{2}\gg T_{1}$ such that the feedback of the first ensemble had effectively locked the LO before the measurement of the second ensemble. In the general setup of locking the LO to $m$ ensembles this corresponds to assuming that $n\gg1$. We will now show how we can apply a phase correction in the measurement of the $j$'th ensemble such that we can relax this assumption. The phase correction will compensate for the fact that the information from the last measurements on the first $(j\!-\!1)$ ensembles has not been fully exploited by the feedback loops before the measurement on the $j$'th ensemble. Note that we assume that the phase correction is only applied to the measurement and not to the LO. 

The phase of the LO relative to the $j$'th ensemble just before the measurement at time $t_{s_{j}n^{j-1}}$ is
\begin{equation}
\Phi^{j}_{s_{j}n^{j-1}}=\sum_{s=1}^{n^{j-1}}\phi_{s+(s_{j}\!-\!1)n^{j-1}}-\Phi^{\text{correct }j}_{s_{j}n^{j-1}},
\end{equation}
where $\phi_{s+(s_{j}\!-\!1)n^{j-1}}=\int_{0}^{T_{1}}\delta\omega(t_{s+(s_{j}\!-\!1)n^{j-1}}-t')\text{d}t'$ and $\Phi^{\text{correct }j}_{s_{j}n^{j-1}}$ is the phase correction applied in the measurement of the $j$'th ensemble at time $t_{s_{j}n^{j-1}}$. Using Eq. \eqref{eq:freqoff}-\eqref{eq:feedback} we can write
\begin{eqnarray}
\Phi^{j}_{s_{j}n^{j-1}}&=&\sum_{s=1}^{n^{j-1}}(\delta\phi_{s+(s_{j}\!-\!1)n^{j-1}}-\alpha\sum_{s'=1}^{s-1}\phi_{s'+(s_{j}\!-\!1)n^{j-1}}^{e_{1}})-\alpha\sum_{s=2}^{n^{j-2}}\sum_{s'=1}^{s-1}\delta\phi^{e_{2}}_{s'n+(s_{j}\!-\!1)n^{j-1}}-\ldots\nonumber \\&&-\alpha\sum_{s=2}^{n}\sum_{s'=1}^{s-1}\delta\phi^{e_{j-1}}_{s'n^{j-1}+(s_{j}\!-\!1)n^{j-1}} -\Phi^{\text{correct }j}_{s_{j}n^{j-1}},
\end{eqnarray}
where $\delta\phi_{s+(s_{j}\!-\!1)n^{j-1}}$ is the accumulated phase between time $t_{(s_{j}-1)n^{j-1}+s-1}$ and $t_{(s_{j}-1)n^{j-1}+s}$ due to the frequency fluctuations of the unlocked LO and the feedback corrections applied up to time $t_{(s_{j}-1)n^{j-1}}$. For simplicity we have replaced the time dependence by an index such that $\delta\phi^{e_{i}}_{s'n^{i-1}+(s_{j}\!-\!1)n^{j-1}}$ is the phase estimate from the $i$'th ensemble at time $t_{s'n^{i-1}+(s_{j}-1)n^{j-1}}$. To fully exploit all information from the measurements on the first $(j\!-\!1)$ ensembles between time $t_{(s_{j}-1)n^{j-1}}$ and $t_{s_{j}n^{j-1}}$,  we choose a phase correction of $\Phi^{\text{correct }j}_{s_{j}n^{j-1}}=\phi^{\text{correct}}_{j,1}+\phi^{\text{correct}}_{j,2}+\ldots+\phi^{\text{correct }}_{j,j-1}$ where
\begin{equation}
\phi^{\text{correct}}_{j,i}=\sum_{s=1}^{n^{j-i}}\left[(1-\alpha)^{n^{j-i}-s}\delta\phi^{e_{i}}_{sn^{i-1}+(s_{j}\!-\!1)n^{j-1}}+\alpha\sum_{s'=1}^{s-1}(1-\alpha)^{n^{j-i}-s}\delta\phi^{e_{i}}_{s'n^{i-1}+(s_{j}\!-\!1)n^{j-1}}\right].
\end{equation}
Here we assume that when two or more ensembles are to be read out at the same instant in time, ensembles with a shorter Ramsey time are measured before the ones with longer Ramsey times such that the results from these measurements can be used as a correction for the ensembles with a longer Ramsey time. For this choice of $\Phi^{\text{correct }j}_{s_{j}n^{j-1}}$ the phase of the LO relative to ensemble $j$ is
\begin{equation} \label{eq:phasej}
\Phi^{j}_{s_{j}n^{j-1}}=\sum_{s=1}^{n^{j-1}}\phi_{s+(s_{j}\!-\!1)n^{j-1}}-\sum_{s=1}^{n^{j-1}}\delta\phi^{e_{1}}_{s+(s_{j}\!-\!1)n^{j-1}}-\sum_{s=1}^{n^{j-2}}\delta\phi^{e_{2}}_{sn+(s_{j}\!-\!1)n^{j-1}}-\ldots-\sum_{s=1}^{n}\delta\phi^{e_{j-1}}_{sn^{j-1}+(s_{j}\!-\!1)n^{j-1}},
\end{equation}
where $\phi_{s+(s_{j}\!-\!1)n^{j-1}}=\int_{0}^{T_{1}}\delta\omega(t_{s+(s_{j}-1)n^{j-1}}-t')\text{d}t'$ is the accumulated phase of the LO relative to the atoms in the first ensemble between times $t_{s-1+(s_{j}-1)n^{j-1}}$ and $t_{s+(s_{j}-1)n^{j-1}}$ . According to Eq. \eqref{eq:phasej}, $\Phi^{j}_{s_{j}n^{j-1}}$ is effectively the accumulated errors between the estimated phases and the actual phases for the $(j-1)$'th ensemble between times $t_{(s_{j}-1)n^{j-1}}$ and $t_{s_{j}n^{j-1}}$ (this is seen by considering Eq. \eqref{eq:phasej} for $j=1,2,\ldots$).  $\Phi^{j}_{s_{j}n^{j-1}}$ is thus the accumulated phase of the LO between time $t_{(s_{j}-1)n^{j-1}}$ and $t_{s_{j}n^{j-1}}$ minus the phase change already measured by the first $j-1$ ensembles, i.e. it does not require further running time to incorporate the information acquired in the first measurements. As opposed to the feedback loop, which  corrects for e.g. frequency drifts by changing the frequency of the LO, the phase corrections directly correct the phase. This phase locking ensures a more rapid convergence, which is important when we want to apply the LO to the subsequent ensembles. With the phase corrections $\Phi^{\text{correct }j}_{s_{j}n^{j-1}}$ we can therefore relax the assumption of $n\gg1$. Since the noise of the LO is white after stabilizing it to the first ensemble the subsequent frequency corrections from the other ensembles could be replaced with merely phase corrections of the LO, which would simplify the above procedure by removing the need for phase corrections in the measurements. We have however chosen to consider frequency corrections to keep a consistent treatment of the feedback in all stages.   
\\

In our simulations we are simulating a clock with a LO locked to $m$ ensembles running for a long but finite time. Similar to our description of the phase corrections $\Phi^{\text{correct }j}_{s_{j}n^{j-1}}$ above there will be some remaining information from the last measurements, which have not been fully exploited by the feedback loops when our simulation stops. In our simulations we therefore include an additional phase correction $\Phi^{\text{correct}}_{\text{final}}$ to the LO after the final measurement. In principle the influence of the last few measurements could also have been reduced by running the simulation for a longer time but by doing the phase correction we reduce the required simulation time. With the phase correction the mean frequency offset of the LO ($\bar{\omega}(\tau)$) after running the clock for a total time of $\tau=lT_{1}$ is
\begin{eqnarray}
\bar{\omega}(\tau)=\frac{1}{\tau}\sum_{s}^{l}\phi_{s}-\Phi^{\text{correct}}_{\text{final}},
\end{eqnarray}
where $\phi_{s}\!=\!\int_{0}^{T_{1}}\delta\omega(t_{s}-t')\text{d}t'$ is the phase of the LO relative to the atoms at time $t_{s}$ and $\Phi^{\text{correct}}_{\text{final}}$ is the final phase correction of the LO. Using Eq. \eqref{eq:freqoff}-\eqref{eq:feedback} and assuming that the $j$'th ensemble is operated with Ramsey time $T_{j}=n^{j-1}T_{1}$ we can write $\bar{\omega}(\tau)$ as:
\begin{equation}
\bar{\omega}(\tau)=\frac{1}{\tau}\left[\sum_{s=1}^{l}(\delta\phi_{s}^{0}-\alpha\sum_{s'=1}^{s-1}\phi_{s'}^{e_{1}})-\alpha\sum_{s=1}^{l/n}\sum_{s'=1}^{s-1}\delta\phi^{e_{2}}_{s'n}-\ldots-\alpha\sum_{s=1}^{l/n^{m-1}}\sum_{s'=1}^{s-1}\delta\phi^{e_{m}}_{s'n^{m-1}}-\Phi^{\text{correct}}_{\text{final}}\right]
\end{equation}
where $\delta\phi_{s}^{0}$ is the accumulated phase between time $t_{s-1}$ and $t_{s}$ due to the frequency fluctuations of the unlocked LO and $\delta\phi^{e_{j}}_{s'n^{j-1}}$ is the estimated phase from the $j$'th ensemble at time $t_{s'n^{j-1}}$. We find that the ideal performance is reached with $\Phi^{\text{correct}}_{\text{final}}=\phi^{\text{correct}}_{\text{final},1}+\phi^{\text{correct}}_{\text{final},2}+\ldots+\phi^{\text{correct}}_{\text{final},m}$ where
\begin{equation}
\phi^{\text{correct}}_{\text{final},j}=\sum_{s=1}^{l/n^{j-1}}\left[(1-\alpha)^{l/n^{j-1}-s}\delta\phi^{e_{j}}_{sn^{j-1}}+\alpha\sum_{s'=1}^{s-1}(1-\alpha)^{l/n^{j-1}-s}\delta\phi^{e_{j}}_{s'n^{j-1}}\right].
\end{equation}
With this phase correction the mean frequency offset is
\begin{equation} \label{eq:meanfreq}
\bar{\omega}(\tau)=\frac{1}{\tau}\left[\sum_{s=1}^{l/n^{m-1}}\tilde{\phi}_{sn^{m-1}}-\delta\tilde{\phi}_{sn^{m-1}}^{e_{1}}-\delta\tilde{\phi}_{sn^{m-1}}^{e_{2}}-\ldots-\delta\tilde{\phi}_{sn^{m-1}}^{e_{m}}\right],
\end{equation} 
where $\tilde{\phi}_{sn^{m-1}}=\sum_{s'=1}^{n^{m-1}}\phi_{(s-1)n^{m-1}+s'}=\sum_{i=1}^{n^{m-1}}\int_{0}^{T_{1}}\delta\omega(t_{s'+(s_{j}-1)n^{m-1}}-t')\text{d}t'$ is the sum of the accumulated phases of the LO relative to the first ensemble between time $t_{s'-1+(s_{j}-1)n^{m-1}}$ and $t_{s'+(s_{j}-1)n^{m-1}}$ and $\delta\tilde{\phi}_{sn^{m-1}}^{e_{j}}=\sum_{s'=1}^{n^{m-j}}\delta\phi^{e_{j}}_{s'n^{j-1}+(s-1)n^{m-j}}$ is the sum of the estimated phases from the $j$'th ensemble at times $t_{s'n^{j-1}+(s-1)n^{m-1}}$. Using Eq. \eqref{eq:phasej} we can write:
\begin{eqnarray}
\bar{\omega}(\tau)&=&\frac{1}{\tau}\left[\sum_{s=1}^{l/n^{m-1}}\tilde{\Phi}^{m}_{s}-\phi_{s}^{e_{m}}\right] \\
&=&\frac{1}{l/n^{m-1}} \left[\sum_{s=1}^{l/n^{m-1}}\frac{\tilde{\Phi}^{m}_{s}-\phi_{s}^{e_{m}}}{T_{m}}\right], \label{eq:meanfreq2}
\end{eqnarray}
where $T_{m}$ is the Ramsey time of the $m$'th ensemble, $\tilde{\Phi}^{m}_{s}$ is the accumulated phase of the stabilized LO relative to the atoms in the $m$'th ensemble at time $t_{sn^{m-1}}$ and $\phi_{s}^{e_{m}}$ is the estimate of that phase. Eq. \eqref{eq:meanfreq2} shows that the final phase correction effectively incorporate the remaining information from the measurements that has not yet been exploited by the feedback loop. Thus the mean frequency offset simply depends on how well we estimate the phase of the $m$'th ensemble and this last measurement is effectively a measurement of the accumulated errors of the phase estimates in the previous $(m-1)$ ensembles. We use Eq. \eqref{eq:meanfreq2} to determine the stability of the clock, which is given by $\sigma_{\gamma}(\tau)=\avg{(\delta\bar{\omega}(\tau)/\omega)^{2}}^{1/2}$.

\section{Modified adaptive measurements}

The adaptive measurement procedure presented in Ref.~\cite{johannesarxiv} can effectively resolve phases between $\pm\pi$ due to the inclusion of rotations of the atomic state. However the assumed dispersive interaction between the probe light and the atoms, which is considered in Ref. \cite{johannesarxiv} would be very challenging to implement for small number of atoms. We have therefore modified the procedure such that the weak measurements are obtained by reading out individual atoms one at a time. Such a procedure is much easier to implement in e.g. ion clocks where individual addressing of ions is feasible. Based on the measurement record we estimate the phase using a Bayesian procedure similar to that of Ref.~\cite{wiseman2000prl,wisemannature2007}. The subsequent feedback on the remaining atoms tries to rotate the atoms into phase with the LO as in Ref.~\cite{johannesarxiv}. 

We will now describe the details of the modified protocol. At the end of the Ramsey sequence, i.e. after the second $\pi/2$ pulse, an atom can either be detected in a spin up state $s=0$ or a spin down state $s=1$. The probability of measuring $s=0,1$ depends on the acquired phase $\delta\phi$ of the LO relative to the atoms during the free evolution in the following way
\begin{equation} 
P(s\vert\delta\phi)=s\cos\left(\pi/4-\delta\phi/2\right)^{2}+(1-s)\sin\left(\pi/4-\delta\phi/2\right)^{2}.
\end{equation}
According to Bayes theorem we can write the probability density of $\delta\phi$ conditioned on the measurement result as
\begin{equation}
P(\delta\phi\vert s)=\frac{P(s\vert\delta\phi)P(\delta\phi)}{P(s)},
\end{equation}
where $P(s)=\int P(\delta\phi)P(s\vert\delta\phi)\text{d}(\delta\phi)$ is the total probability of measuring $s$ and $P(\delta\phi)$ is the a priori probability distribution of $\delta\phi$, which is determined from characterizing the frequency fluctuations of the LO. We choose a Gaussian distribution with zero mean and variance $\gamma T$ as the a priori distribution. This a priori distribution is exact for a LO subject to white noise but a better a priori distribution could possibly be found for $1/f$ noise. We will however use this a priori distribution in both cases, which results in our modified protocol not being as effective for $1/f$ noise as for white noise in the LO. Note that this inaccuracy in our a priori distribution only introduce a less ideal performance in our phase estimate. In our numerical simulations we retain the full information about the phase evolution so that our suboptimal assumption about the a priori distribution only degrade the performance of the scheme. 
We estimate the phase based on the measurement as
\begin{equation}
\delta\phi^{e}=\int \delta\phi P(\delta\phi\vert s)\text{d}(\delta\phi).
\end{equation}
We then apply a feedback to the remaining atoms, which in the Block sphere picture rotates them by an angle $\delta\phi^{e}$ around the $\hat{J}_{x}$ axis, i.e. the feedback tries to bring them into phase with the LO. Generalizing this procedure to a measurement record $S_{m}=s_{1}s_{2}\ldots s_{i} \ldots s_{m}$ where the measurement result of the $i$'th atom is $s_{i}$, we obtain
\begin{equation}
P(S_{m}\vert \delta\phi,\sum_{k=1}^{n_{m}}\delta\phi^{e}_{k}) = \prod_{i}^{m}\left[s_{i}\cos\left(\pi/4-\delta\phi/2+\sum_{k=1}^{n_{i}}\delta\phi^{e}_{k}/2\right)^{2}+(1-s_{i})\sin\left(\pi/4-\delta\phi/2+\sum_{k=1}^{n_{i}}\delta\phi^{e}_{k}/2\right)^{2}\right],
\end{equation}  
where $P(S_{m}\vert \delta\phi,\sum_{k=1}^{n_{m}}\delta\phi^{e}_{k})$ is the probability of obtaining the measurement record $S_{m}$ conditioned on a drifted phase $\delta\phi$ with a total feedback of $\sum_{k=1}^{n_{m}}\delta\phi^{e}_{k}$ applied during the measurements ($\sum_{k=1}^{n_{i}}\delta\phi^{e}_{k}$ is the feedback experienced by the $i'th$ atom before it is read out). Note that in general $n_{i}\neq i-1$, i.e. we might read out more than one atom before we do a phase estimate and a subsequent feedback on the remaining atoms. In our simulations we group the measurements such that we perform $\sim4$ feedbacks in total as in the protocol of Ref.~\cite{johannesarxiv} for uncorrelated atoms. The final phase estimate $\delta\phi^{e}_{n_{m}+1}$ after having read out $m$ atoms is
\begin{equation}
\delta\phi^{e}_{n_{m}+1}=\frac{\int (\delta\phi-\sum_{k=1}^{n_{m}}\delta\phi^{e}_{k})P(S_{m}\vert\delta\phi,\sum_{k=1}^{n_{m}}\delta\phi^{e}_{k})P(\delta\phi)\text{d}(\delta\phi)}{\int P(S_{m}\vert\delta\phi,\sum_{k=1}^{n_{m}}\delta\phi^{e}_{k})P(\delta\phi)\text{d}(\delta\phi)},
\end{equation}
where we have used Bayes theorem as described above. All the phase estimates and the feedbacks are performed after the final $\pi/2$ pulse in the Ramsey sequence. Thus the final estimate of the drifted phase is $\delta\phi_{e}=\sum_{k=1}^{n_{m}+1}\delta\phi^{e}_{k}$, i.e. the sum of the rotations performed during the measurements and the final phase estimate.

\section{Limit of the free evolution time} 

As described in the beginning of the main article the stability of an atomic clock increases with the Ramsey time $T$. For clocks with trapped atoms $T$ is essentially only limited by the decoherence in the system, which in practice often originates from the LO. This decoherence is what results in the phase offset between the LO and the atoms after the period of free evolution in the Ramsey sequence. For experiments and simulations running with fixed Ramsey times there is a finite probability that a phase jump occurs, which is large enough to spoil the feedback strategy used to lock the LO to the atomic transition. This can result in the feedback jumping to a state with a phase difference of $2\pi$ (so called \emph{fringe hops} \cite{rosenband2012arxiv}) or the measurement strategy can fail leading to ambiguous results. For the conventional Ramsey protocol this happens for phase jumps larger than $\pi/2$ while the adaptive protocol breaks down for phase jumps larger than $\pi$ \cite{johannesarxiv}. The probability of these phase jumps increases with $T$ since the width of the distribution of the phase jumps $\sigma^{2}_{\delta\phi}$ increases with $T$, e.g. for white noise $\sigma^{2}_{\delta\phi}=\gamma T$. One could include correction strategies for the errors introduced by these phase jumps (e.g. running an ensemble with different Ramsey times would correct for the fringe hops) but for simplicity we do not consider this in our simulations. 

As shown in the main article the minimum number of atoms required in each ensemble in order to increase the Ramsey time by a factor of $a$ at each level of our protocol is 
\begin{equation} \label{eq:nmin}
N\sim a/\beta
\end{equation}
where $\beta$ parameterize the maximal Ramsey time $T_{max}$ for a LO subject to white noise.  Note that equivivalently $\beta$ is the maximal width of the distribution of phase jumps allowed, i.e. $\sigma^{2}_{\delta\phi,max}=\beta$. The requirement expressed in Eq. \eqref{eq:nmin} ensures that when we increase the Ramsey time of the next ensemble by a factor of $a$ compared to the Ramsey time of the previous ensemble we still keep $\sigma^{2}_{\delta\phi} \lesssim \sigma^{2}_{\delta\phi,max}$ for the noise seen by the next ensemble. Note that minimum number of atoms required for our protocol to work is found by setting $a=2$.  

\begin{figure} [H]
\centering
\subfloat {\label{fig:ramseytime_white}\includegraphics[width=0.4\textwidth]{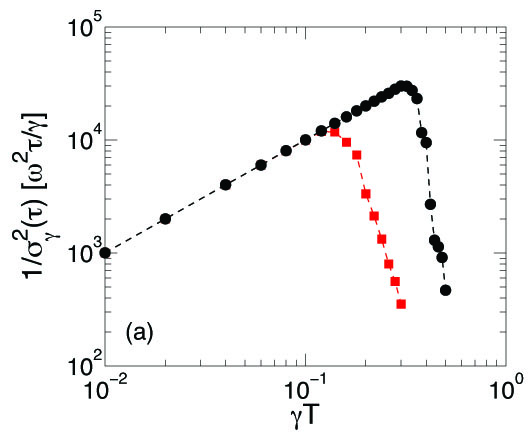}} 
\subfloat{\label{fig:ramseytime_pink}\includegraphics[width=0.4\textwidth]{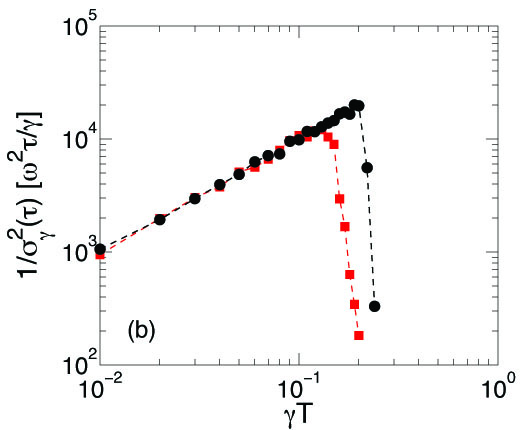}}
\caption{Stability as a function of the Ramsey time ($\gamma T$) for (a) white noise and (b) $1/f$ noise in the LO. The plots were made with $N=10^{5}$. $\bullet$ is the adaptive protocol of Ref.~\cite{johannesarxiv} and $\blacksquare$ is the conventional Ramsey protocol. The adaptive protocol allows for $\gamma T \sim 0.3$ and $0.2$ for white and $1/f$ - noise respectively while the conventional protocol only allows for $\gamma T \sim 0.1$ for both white and $1/f$ noise.}
\label{fig:breakdown}
\end{figure}

To determine $\beta$ we simulate an atomic clock with only a single ensemble with $N=10^{5}$ atoms and a LO subject to white noise characterized by a strength $\gamma$. Furthermore to determine $\beta_{1}$ (see Eq. (11) and above in the article) for a LO subject to $1/f$ noise we do a similar simulation but with a $1/f$ noise spectrum of the LO i.e. $S(f)=\gamma^{2}/f$ where $S(f)$ denotes the noise spectrum and $f$ is frequency. Note that we define the noise spectrum as $S(f)\delta(f+f')=\avg{\delta\omega(f)\delta\omega(f')}$ where $\delta\omega(f)$ is the Fourier transform of the frequency fluctuations $\delta\omega(t)$ of the LO. $S(f)$ is thus the frequency noise spectrum. In the simulations we do not simulate the full quantum evolution of the atomic state as we do for the simulations presented in the article. Instead we approximate the probability distributions of $\hat{J}_{x,y,z}$ with Gaussian distributions as in Ref.~\cite{johannesarxiv}. This Gaussian approximation is legitimate since $N\gg1$. Furthermore it is desirable to have a weak feedback strength $\alpha$ for white noise in the unlocked LO since a strong feedback increases the width of the phase noise for the locked LO. For white noise in the unlocked LO we therefore simulate the limit where  $\alpha\ll1$ such that the phases are uncorrelated. For $1/f$ noise we use a feedback strength of $\alpha=0.5$ since a stronger feedback is desirable to lock the LO more rapidly. The high number of atoms ensures that when we increase the Ramsey time $T$ of the clock we see the onset of the phase jumps as an abrupt break down, which is not blurred by the atomic noise in our phase estimates. In our simulation the clock is running for a time $\tau=10^{6}T$, i.e. for $l=10^{6}$ steps of $T$ (for $1/f$ noise we average over 100 independent runs with $10^{4}$ steps of $T$). The onset of the break down will in principle have a weak (logarithmic) dependence on the number of steps that we simulate, which we do not expect to change our results significantly \cite{johannesarxiv}. \figref{fig:breakdown} shows the result of our simulations. 
The adaptive protocol that we have used here is that of Ref.~\cite{johannesarxiv} since the modified adaptive protocol will lead to similar results for large atom numbers where the break down is most apparent but is harder to simulate.  
\figref{fig:breakdown} shows that the conventional protocol allows for $\beta\sim0.1$ for white noise and that $\beta_{1}\sim0.1$ for $1/f$ noise in the unlocked LO while the adaptive protocol allows for $\beta\sim0.3$ and $\beta_{1}\sim0.2$ for $1/f$ noise. With $a=2$ in Eq. \eqref{eq:nmin} the minimum number of atoms required for the protocol of locking to several ensembles to work is thus $N_{min}=20$ for conventional Ramsey strategy while the adaptive strategy can extend the applicability down to $N_{min}=7$ atoms. We expect $\beta$ of the modified adaptive protocol to be identical to that of the adaptive protocol in Ref.~\cite{johannesarxiv} since both rely on the rotation of the atomic state to resolve phases between $\pm\pi$. In our numerical simulations of the modified protocol we have therefore set $\beta\sim0.3$ and $\beta_{1}\sim0.2$ for $1/f$ noise. Note that in our simulations of the full protocol of locking an atomic clock to several atomic ensembles, we still include the possibility of disruptive phase jumps. However imposing the limits on $\beta$ ($\beta_{1}$) identified from \figref{fig:breakdown} for all steps in the protocol ensures that we do not see any significant effect of them. The probability to have disruptive phase jumps for the duration of the simulations is simply negligible, i.e, the probability for phase jumps large enough to spoil the feedback strategy in a Ramsey sequence is well below $10^{6}$. Note that the feedback strength is set to $\alpha=0.01$ for white noise and $\alpha=0.5$ for $1/f$ noise in the LO in our simulations of the full protocol. The strong feedback strength of $\alpha=0.5$ is only used for the first ensemble for $1/f$ noise since the noise seen by the other ensembles is white and a weaker feedback strength is thus desirable.     

In our above estimates of $N_{min}$ we have assumed that the adaptive protocol leads to a stability at the SQL. This is only true for large N and there are corrections to this for small $N$.  From our simulations we find that with the modified adaptive protocol, and white noise in the unlocked LO, the feedback of the first ensemble stabilizes the LO to a white noise floor below $1/NT_{1,max}$ (see Eq. (8)), i.e. better than what we expect from the SQL. We can thus extend the applicability of the protocol to atom numbers below $N=7$. As shown in Fig. 3 in the main article we find that we can go as low as 4 atoms and still have the feedback of first ensemble lowering the noise floor of the LO by a factor of two such that the second ensemble can be operated with twice as long a Ramsey time while keeping the width of the phase noise below $\beta=0.3$. For $1/f$ noise the minimum number of atoms is still $N=7$. This slightly worse performance is a result of the shorter required Ramsey time for the first ensemble ($\beta_{1}\sim0.2$) and the incomplete characterization of the a priori probability distribution in the Bayesian approach (see the section "Modified adaptive measurements" above). The latter results in the feedback only stabilizing the LO to the SQL of  $\sim1/NT_{1,max}$ also for small $N$ and the minimum number of atoms is thus 7 as seen from Eq. \eqref{eq:nmin} with $a=2$. Note, however, that in principle it is only in the first ensemble that we need more atoms than for white noise since the feedback of the first ensemble will have whitened the noise affecting the subsequent ensembles. The modified adaptive protocol is thus more effective for the subsequent ensembles.

\end{document}